\documentclass[a4paper]{PoS}

\newcommand{\apj}{ApJ}
\newcommand{\Ds}{{D_{\rm s}}}
\newcommand{\Dd}{{D_{\rm d}}}
\newcommand{\Dds}{{D_{\rm ds}}}
\newcommand{\rE}{{r_{\rm E}}}
\newcommand{\tE}{{t_{\rm E}}}
\newcommand{\vt}{{v_{\rm t}}}
\newcommand{\murel}{{\mu_{\rm rel}}}
\newcommand{\thetaE}{{\theta_{\rm E}}}
\newcommand{\ys}{{y_{\rm s}}}
\newcommand{\xs}{{x_{\rm s}}}
\newcommand{\rs}{{r_{\rm s}}}
\newcommand{\zs}{{z_{\rm s}}}
\newcommand{\fs}{{f_{\rm s}}}
\newcommand{\ri}{{r}}
\def\yr{\,{\rm yr}}
\def\day{\,{\rm day}}
\def\cm{\,{\rm cm}}
\def\kpc{\,{\rm kpc}}
\def\kms{\,{\rm km\,s^{-1}}}

\def\AU{\,{\rm AU}}
\def\mas{\,{\rm mas}}
\def\rhos{\rho_\star}
\def\zbar{\bar{z}}
\newcommand{\beq}{\begin{equation}}
\newcommand{\eeq}{\end{equation}}

\title{Introduction to Gravitational Microlensing}

\ShortTitle{Introduction to Gravitational Microlensing}

\author{\speaker{Shude Mao} \\
        Jodrell Bank Centre for Astrophysics, University 
	of Manchester, Manchester M13 9PL, UK \\
        E-mail: \email{shude.mao@manchester.ac.uk}}

\abstract{
The basic concepts of gravitational microlensing are introduced.
We start with the lens equation, and then derive 
the image positions and magnifications. The statistical quantities of optical
depth and event rate are then described. We finish with a summary and a
list of challenges and open questions. A problem set is given for students to practice.
}

\FullConference{The Manchester Microlensing Conference: The 12th International Conference and ANGLES Microlensing Workshop\\
		 January 21-25 2008\\
		 Manchester, UK}
\begin{document}

\section{Introduction}

Gravitational microlensing (in the local group) refers to the temporal
brightening of a background star due to intervening objects.
Einstein (1936) first examined (micro)lensing by a single
star, and concluded that ``there is no great chance of observing this phenomenon.''
Some important works were performed in intervening years by \cite{Ref64}
and \cite{Lie64}, but the research topic was revitalised by Paczy\'nski (1986) who proposed it as a method
to detect compact dark matter objects in the Galactic halo.

The original goal is now out of favour, since we know with
high precision that most of the dark matter must be non-baryonic,
e.g. from observations of microwave background radiation and
nucleosynthesis (at the time of his paper, this was, however, unclear).
Nevertheless, gravitational microlensing has turned into a powerful
technique with diverse applications in
astrophysics, including the study of the structure of the Milky Way,
stellar atmospheres and the detection of extrasolar planets and
stellar-mass black hole candidates. The field has made enormous progress in the last two decades. There
have been a number of reviews on this topic (e.g. \cite{Pac96, Mao01, Eva03, Wam06}), the
most recent highlight was given in \cite{Gou08a}.
This article gives an introduction to microlensing, aimed at a level for a starting PhD student. 
Together with other talks in the workshop and proceedings 
\footnote{available at {\tt http://pos.sissa.it/cgi-bin/reader/conf.cgi?confid=54}}, one can gain a thorough feeling
about the state-of-the-art research in this field (as of 2008).

The reference list given here is seriously incomplete (and biased). For more complete
references and information about ongoing microlensing surveys,
see the review papers mentioned above and the 
web site:  {\tt http://mlens.net/} (built by Szymon Koz{\l}owski, Subo Dong and Lukasz Wyrzykowski).

\section{What is gravitational microlensing?}

\begin{figure}[!ht]
\includegraphics[width=.45\textwidth]{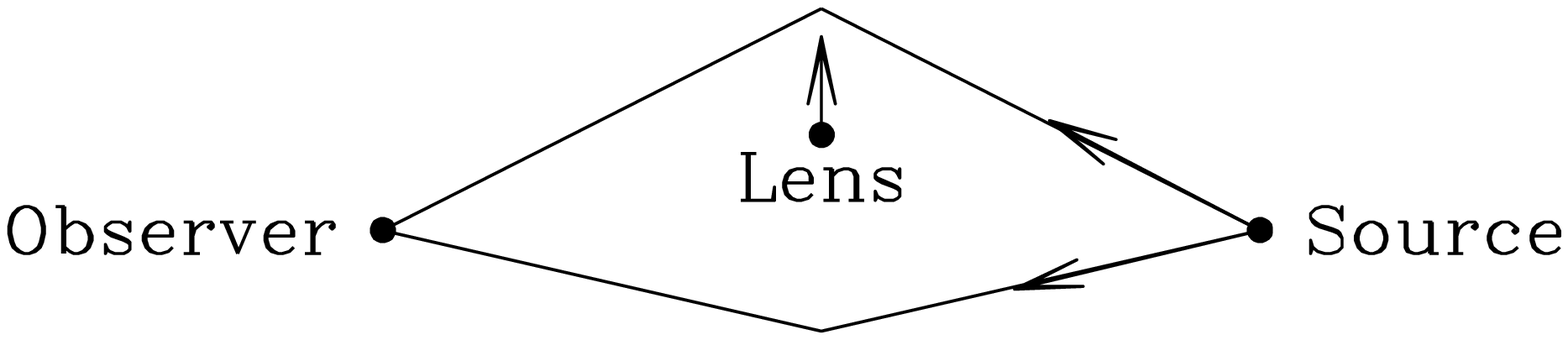} \hfil
\includegraphics[width=.45\textwidth]{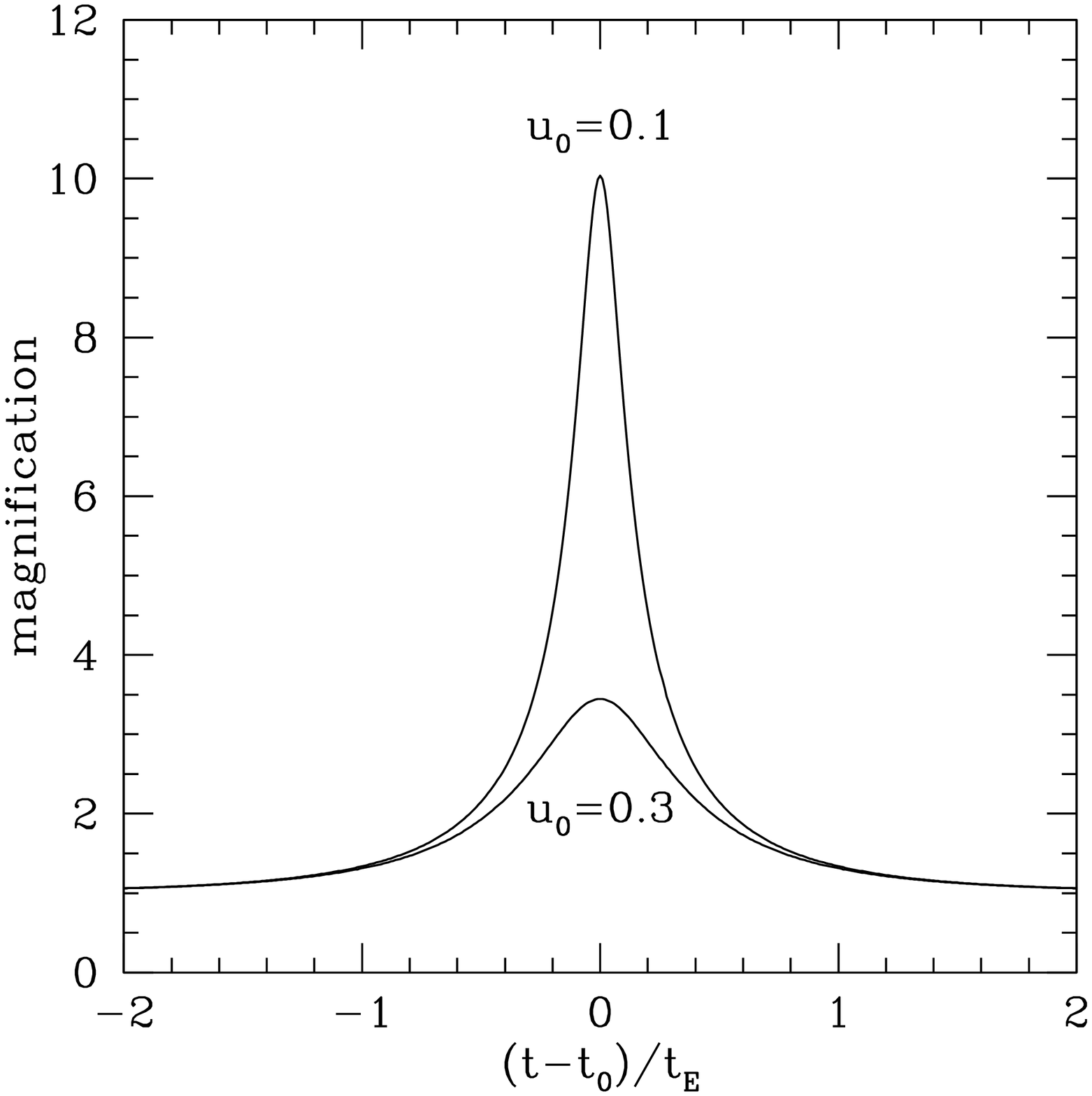}
\caption{The left panel shows a side-on view of the geometry of microlensing
  where a lens moves across the line of sight towards a background source. The
  right panel shows two light curves corresponding to two dimensionless impact
  parameters, $u_0=0.1$ and 0.3. The time on the horizontal axis is centred on the
peak time $t_0$ and is normalised to the Einstein radius crossing time
$\tE$. The lower the value of $u_0$, the higher the peak
magnification. For the definitions of $u_0$ and $\tE$ see section \protect\ref{sec:standard}.}
\label{fig:lc}
\end{figure}

The light from a background source is deflected, distorted and
(de)magnified by intervening objects along the line of sight. If the
lens, source and observer are sufficiently well aligned, then strong gravitational
lensing can occur. Depending on the lensing object, strong gravitational
lensing can be divided into three areas: microlensing by stars,
multiple-images by galaxies, and giant arcs and large-separation lenses
by clusters of galaxies. For microlensing, the lensing object is a stellar-mass compact object (e.g. normal
stars, brown dwarfs or stellar remnants [white dwarfs, neutron stars and
  black holes]); the image splitting in this case is
usually too small (of the order of milli-arcsecond in the local group) to be resolved by
ground-based telescopes, thus we can only observe the magnification change as a function of time.

The left panel in Fig.  \ref{fig:lc} illustrates the
microlensing geometry. A stellar-mass lens moves across the line of sight towards a
background star. As the lens moves closer to the line of sight, its gravitational
focusing increases, and the background star becomes brighter. As the source moves
away, the star falls back to its baseline brightness. If the motions of
the lens, the observer and the source can be approximately taken as
linear, then the light curve is symmetric. Since the lensing probability
for microlensing in the local group is of the order of $10^{-6}$
(see section \ref{sec:tau}), the microlensing variability usually should not
repeat. Since photons of different wavelengths follow the same propagation path
(geodesics), the light curve (for a point source) should
not depend on the colour. The characteristic symmetric shape,
non-repeatability, and achromaticity can be used as criteria to separate microlensing
from other types of variable stars (exceptions to these rules will be
discussed in section \ref{sec:exotic}).

\section{Lens equation, image positions and magnifications}

To derive the characteristic light curve shape shown in the right panel
of Fig. \ref{fig:lc}, we must look closely at the lens equation, and the resulting image positions and magnifications for a point source.

\subsection{Lens equation}

\begin{figure}[!ht]
\begin{center}
\includegraphics[width=0.9\textwidth]{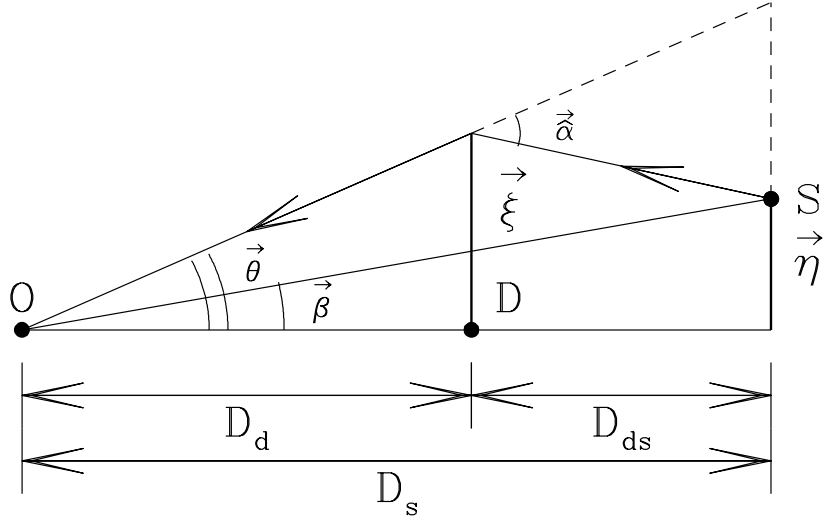}
\end{center}
\vspace{-6.5cm}
\caption{Illustration of various distances and angles in the lens 
equation (eqs. \protect\ref{eq:lensPhysical} and \protect\ref{eq:vectorlens}).
}
\label{fig:lenseq}
\end{figure}

The lens equation is straightforward to derive. Figure 
\ref{fig:lenseq} illustrates a side-on view of the lensing
configuration. Simple geometry yields
\beq
\label{eq:lensPhysical}
\vec{\eta} + \Dds \hat{\vec{\alpha}} = \vec{\xi} \cdot \frac{\Ds}{\Dd},
\eeq
where $\Dd$, $\Ds$ and $\Dds$ are the distance to the lens (deflector), distance to the
source and distance between the lens (deflector) and the source,
$\vec{\eta}$ is the source position (distance perpendicular to the line
connecting the observer and the lens), $\vec{\xi}$ is the image position,
and $\hat{\vec{\alpha}}$ is the deflection angle. For gravitational microlensing in
the local group, $\Dds=\Ds-\Dd$.\footnote{For cosmological microlensing
  in an expanding universe,
  the distances should be taken as angular diameter distances, and in
  general $\Dds \ne \Ds-\Dd$. See the review by Wambsganss in these
  proceedings on cosmological microlensing.}
Mathematically, the lens equation provides a mapping between the source plane to the lens plane. The mapping 
is not necessarily one-to-one.

Dividing both sides of eq. (\ref{eq:lensPhysical}) by $\Ds$, we obtain
the lens equation in angles
\beq
\vec{\beta} + \vec{\alpha} = \vec{\theta},
\label{eq:vectorlens}
\eeq
where $\vec{\beta}=\vec{\eta}/\Ds$, $\vec{\theta}=\vec{\xi}/\Dd$, and
$\vec{\alpha} = \hat{\vec{\alpha}} \times \Dds/\Ds$ is the scaled
deflection angle. These angles are illustrated in Fig. \ref{fig:lenseq}.

For an axis-symmetric mass distribution, due to symmetry, the source, observer and image positions
must lie in the same plane, and so we can drop the vector sign, and
obtain a scalar lens equation:
\beq
{\beta} + {\alpha} = {\theta}.
\eeq

\subsection{Image positions for a point lens}

For a point lens at the origin, the deflection angle is given by
\beq
\hat{\vec{\alpha}} = \frac{4 G M}{c^2} \frac{1}{\xi^2} \vec{\xi},
\eeq
and the value of the scaled deflection angle is
\beq
\alpha = \frac{\Dds}{\Ds} |\hat{\vec{\alpha}}|=
\frac{\Dds}{\Ds} \frac{4 G M}{c^2 \Dd \theta} \equiv \frac{\thetaE^2}{\theta}, ~~~ \xi=\Dd \theta.
\eeq
where we have defined the angular Einstein radius as
\beq
\thetaE = \frac{\rE}{\Dd} \approx
0.55 \mas \sqrt{\frac{1-\Dd/\Ds}{\Dd/Ds}} \left(\frac{\Ds}{8 \kpc}\right)^{-1/2}\left(\frac{M}{0.3 M_\odot}\right)^{1/2}.
\label{eq:thetaE}
\eeq
The lens equation for a point lens in angles is therefore
\beq
\beta + \frac{\thetaE^2}{\theta} = \theta.
\eeq
We can further simply by normalising all the angles by $\thetaE$,
$\rs \equiv \beta/\thetaE$, $\ri \equiv \theta/\thetaE$, the above
equation becomes\footnote{$\rs$ is not to be confused with the size of the
  star, which we denote as $r_\star$.}
\beq
\rs  + \frac{1}{\ri} = \ri.
\label{eq:rs}
\eeq
For the special case when the lens, source and observer are perfectly
aligned ($\rs=0$), due to axis-symmetry along the line of sight, the
images form a ring (``Einstein'' ring) with its angular size given by eq. (\ref{eq:thetaE}).

For any other source position $\rs \ne 0$, there are always two images, their
positions are given by
\beq
\ri_{\pm} = \frac{\rs \pm \sqrt{\rs^2+4}}{2}.
\eeq
The `+' image is outside the Einstein radius ($\ri_{+} \ge 1$) on the same
side of the source, while the `$-$' image is on the opposite side and
inside the Einstein radius ($\ri_{-}<0$ and $|\ri_{-}|<1$). The angular separation between
the two images is
\beq
\Delta \theta = \thetaE (\ri_{+} - \ri_{-}) = \thetaE \sqrt{4+\rs^2}.
\eeq
The image separation is of the same order of the angular Einstein diameter when
$\rs \lesssim 1$, and thus will be in general too small to be observable given the typical
seeing from the ground ($\sim$ one arcsecond); we can only observe
lensing effects through magnification. One exception may be the VLT interferometer (VLTI)
which can potentially resolve the two images. This may be important for
discovering stellar-mass black holes since they have larger image
separations due to their larger masses than typical lenses with mass
$\sim 0.3 M_\odot$ (\cite{Del01, RM06}).

The physical size of the Einstein radius in the lens plane is given by
\begin{equation}
\rE = \Dd\thetaE = \sqrt{ \frac{4 GM}{c^2} \frac{\Dd \Dds}{\Ds}}
\approx 2.2 \AU \sqrt{4 \times \frac{\Dd}{\Ds}\left(1-\frac{\Dd}{\Ds}\right)} \left(\frac{\Ds}{8 \kpc}\right)^{1/2}  \left(\frac{M}{0.3 M_\odot}\right)^{1/2}. 
\label{eq:rE}
\end{equation}
So the size of the Einstein ring is roughly the scale of the solar
system, which is a coincidence that helps the discovery of extrasolar
planets around lenses.

\subsection{Image magnifications}

\begin{figure}[!h]
\begin{center}
\includegraphics[width=0.7\textwidth]{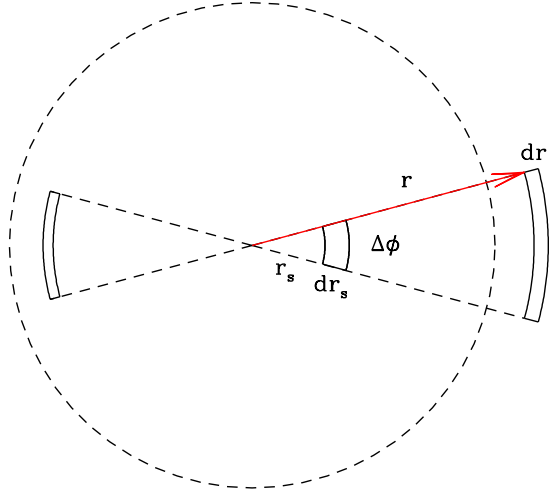}
\end{center}
\vspace{-3cm}
\caption{Images of a thin annulus from $\rs$ to $\rs+d\rs$ by a point
lens on the plane of the sky. The dashed line is the Einstein
ring. $\Delta\phi$ is the angle subtended by the thin annulus.
}
\label{fig:mag}
\end{figure}

Since gravitational lensing conserves surface brightness, the
magnification of an image is simply given by the ratio of the image area
and source area. For a very small source, we can consider a thin source
annulus with angle $\Delta \phi$ (see Fig. \ref{fig:mag}). For a point lens, this
thin annulus will be mapped into two annuli, one inside the Einstein ring and one outside.

The area of the source annulus is given by the product of the radial
width and the tangential length $d\rs \times \rs\Delta\phi$. Similarly, 
each image area is $d\ri \times \ri\Delta\phi$, and the
magnification is given by\footnote{This is a special case of the
  determinant of the Jacobian in the lens mapping, see section \protect\ref{sec:binary}.}
\beq
\mu = \frac{ d\ri \times \ri\Delta\phi}{d\rs \times \rs\Delta\phi} 
= \frac{\ri}{\rs} \frac{d\ri}{d\rs}.
\label{eq:image}
\eeq
For the two images given in eq. (\ref{eq:image}), one finds
\beq
\mu_{+} = \frac{(\rs+\sqrt{\rs^2+4})^2}{4 \rs \sqrt{\rs^2+4}}, ~~
\mu_{-} = -\frac{(\rs-\sqrt{\rs^2+4})^2}{4 \rs \sqrt{\rs^2+4}}.
\eeq
The magnification of the `+' image is positive, while the `$-$' image is
negative. The former image is said to have positive parity while the
latter negative\footnote{Let us imagine two arrows for the thin
annulus (see Fig. \ref{fig:mag}), one in the radial direction
and one in the tangential direction respectively. For the negative parity image, the
corresponding tangential arrow for the image is reversed with respect to
that in the source, while in the radial direction the arrow directions
remain the same for the source and image. For the positive
parity image, the directions of the arrows are the same for the image
and the source.}. The total magnification is given by
\beq
\mu = |\mu_{+}| +|\mu_{-}| = \frac{\rs^2+2}{\rs\sqrt{\rs^2+4}},
\label{eq:mu}
\eeq
and the difference is identical to unity
\beq
|\mu_{+}|-|\mu_{-}|=1.
\eeq
We make some remarks about the total magnification and image separations:
\begin{enumerate}
\item When $\rs=1$, $\mu=3/\sqrt{5}\approx 1.342$,  corresponding to about 0.319
  magnitude. Such a difference is easily observable\footnote{For bright
stars, the accuracy of photometry can reach a few milli-magnitudes.}, and so the area
  occupied by the Einstein ring is usually taken as the lensing ``cross-section.''
\item When $\rs \rightarrow \infty$, $|\mu_{+}/\mu_{-}| \rightarrow
  \rs^{4}$, $\mu \rightarrow 1+2 \rs^{-4}$. The angular image
  separation is given by $\Delta\theta = (\rs+2 \rs^{-1})\thetaE$.
\item High magnification events occur when $\rs \rightarrow 0$. 
The asymptotic behaviours are $\mu \rightarrow \rs^{-1}(1+3 \rs^2/8)$, $\Delta\theta
\rightarrow (2+\rs^2/4)\thetaE$, and $dr/d\rs \rightarrow 1/2$. The last
relation implies that, at high magnification, the image is compressed by a
factor of 2 in the radial direction (see Fig. \ref{fig:mag}).
\end{enumerate}

\vspace{-2cm}

\section{Light curve and microlensing degeneracy}

Given a source trajectory, we can easily describe the standard light curve with
a few parameters which suffers from the microlensing degeneracy.

\subsection{Standard light curve \label{sec:standard}}

\begin{figure}[!h]
\begin{center}
\includegraphics[width=0.7\textwidth]{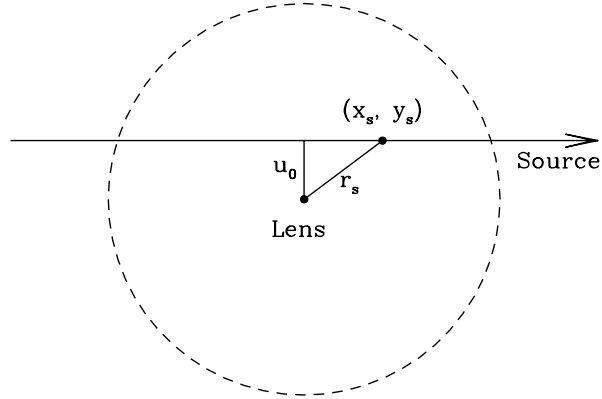}
\end{center}
\vspace{-3cm}
\caption{Illustration for the lens position and source trajectory. The
  dimensionless impact parameter is $u_0$,
$(\xs, \ys)$ are the dimensionless source position along the trajectory,
and $\rs$ is the distance between the lens and source.}
\label{fig:traj}
\end{figure}

For convenience, we put the lens at the origin, and let the source move across the line
of sight along the $x$-axis (see Fig. \ref{fig:traj}). The impact parameter
in units of the Einstein radius is labelled as $u_0$. For convenience,
we define the Einstein radius crossing time (or `timescale') as
\beq
\tE = \frac{\rE}{\vt} = \frac{\thetaE}{\murel}, ~~
\thetaE=\frac{\rE}{\Dd}, ~~ \murel \equiv \frac{\vt}{\Dd}
\label{eq:tE}
\eeq
where $\vt$ is the transverse velocity and $\murel$ is the relative
lens-source proper motion. Substituting the expression for the Einstein
radius into eq. (\ref{eq:rE}), we find that
\beq
\tE \approx 19\,\day \sqrt{4\times \frac{\Dd}{\Ds}\left(1-\frac{\Dd}{\Ds}\right)} \left(\frac{\Ds}{8 \kpc}\right)^{1/2} 
\left(\frac{M}{0.3M_\odot}\right)^{1/2}
\left(\frac{\vt}{200\kms}\right)^{-1}.
\eeq
If the closest approach is achieved a time $t=t_0$, then the
dimensionless coordinates are $\xs = (t-t_0)/\tE$ and $\ys = u_0$, and the
magnification as a function of time is given by
\beq
\mu(t) = \frac{\rs^2(t)+2}{\rs(t)\sqrt{\rs^2(t)+4}}, ~~~ \rs(t) =
\sqrt{u_0^2 + \left(\frac{t-t_0}{\tE}\right)^2}.
\label{eq:standard}
\eeq
Two light curve examples are shown in the right panel of Fig. \ref{fig:lc} for $u_0=0.1$ and 0.3.

\vspace{-2cm}

To model an observed light curve, three parameters are present in eq. (\ref{eq:standard}): $t_0$,
$\tE$, $u_0$. In practise, we need two additional parameters, $m_0$, the
baseline magnitude, and $\fs$, a blending parameter. $\fs$ 
characterises the fraction of light contributed by the lensed source; in
crowded stellar fields, each observed `star' may be a composite of the
lensed star, other unrelated stars within the seeing disk and the lens
if it is luminous (\cite{Alc01b, Koz07}). Blending will lower the observed magnification and in general $\fs$ depends on the
wavelength, and so each filter requires a separate $\fs$ parameter.
Unfortunately, we can see from eq. (\ref{eq:standard}) that
there is only one physical parameter ($\tE$) in the model that relates to the lens properties.
$\tE$ depends on the lens mass, distances to the lens and source, and
the transverse velocity $\murel$.  Therefore from an observed light
curve well fitted by the standard model, one cannot infer the
lens distance and mass uniquely; this is the so-called microlensing
degeneracy. However, given a lens mass function and some kinematic model of the
Milky Way, we can infer the lens mass statistically.

\subsection{Non-standard light curves \label{sec:exotic}}

The standard model assumes the lensed source is point-like, both the
lens and source are single and all the motions are linear.
The majority ($\sim 90\%$) of microlensing events are well described by
this simple model. However, about 10\% of the light curves are
non-standard (exotic), due to  the breakdown of one (or more) of the assumptions.
We briefly list these possibilities below (see the talk by Dominik for more details.)
These non-standard microlensing events allow us to derive extra constraints, and partially
lift the microlensing degeneracy. Because of this, they play a role far
greater than their numbers suggest.
\begin{enumerate}
\item[(1)] The lens may be in a binary or even a multiple system
(\cite{MP91}). The light curves for a binary or multiple lensing system
can be much more diverse (see \ref{sec:binary}). They offer an exciting
way to discover extrasolar planets (\cite{MP91, GL92, BR96, GS98, Rat02}).
\item[(2)] The source is in a binary. In this case, the light curve will
  be a simple, linear superposition of the two sources (when the rotation can be
  neglected, see \cite{GH92}).
\item[(3)] The finite size of the lensed star cannot be neglected. This
occurs when the impact parameter $u_0$ is comparable to the stellar
radius normalised to the Einstein radius, $u_0 \sim r_\star/\rE$. In
this case, the light curve is significantly modified by the finite
source size effect (\cite{WM94, Gou94}). The finite source size effect
is most important for high magnification events.
\item[(4)] The standard light curve assumes all the motions are
linear. However, the source and/or the lens may be in a binary,
furthermore, the Earth rotates around the Sun. All these motions induce
accelerations. The effect due to the Earth motion around the Sun is
usually called ``parallax'' (e.g. \cite{Gou92, Smi02, Poi05}), while that due to binary motion in the
source plane is called ``xallarap'' (``parallax'' spelt backwards,
\cite{Ben98, Alc01}). Parallax or ``xallarap'' events usually have long
timescales. For a typical microlensing event with timescale $\tE \sim
20\day$, the parallax effect due to the Earth rotation around the Sun is often undetectable (unless
the photometric accuracy of the light curve is very high).
\item[(5)] Microlensing can ``repeat'', in particular if the lens is a wide
  binary (\cite{Dis96}) or the source is a wide binary. In such cases, microlensing may
  manifest as two well-separated peaks, i.e., as a ``repeating'' event. A
  few percent of microlensing events are predicted to repeat, consistent
  with the observations (\cite{Sko08}).
\end{enumerate}
Notice that several violations may occur at the same time, which in some
cases allow the microlensing degeneracy to be completely removed
(e.g. \cite{An02, Don08, Gau08}).

\subsection{$N$-point lens gravitational microlensing \label{sec:binary}}

It is straightforward to derive the (dimensionless) lens equation for $N$-point lenses. We can first cast
eq. (\ref{eq:rs}) in vector form and then rearrange
\beq
\vec{\rs} = \vec{\ri} - \frac{1}{|\vec{\ri}|^2} \vec{\ri}.
\label{eq:onelens}
\eeq
The above expression implicitly assumes that the lens is at the origin, and all the lengths have been normalised to the Einstein radius
corresponding to its mass (or equivalently, the lens mass has been assumed to be unity).

Let us consider the general case where we have $N$-point lenses, at
$\vec{r}_k=(x_k, y_k)$ with mass $M_k$, $k=1, \cdot\cdot\cdot, N$. We normalise
all the lengths with the Einstein radius corresponding to the total
mass, $M=\sum_{k=1}^N M_k$. The generalised lens equation then reads
\beq
\vec{\rs} = \vec{\ri} - \sum_{k=1}^{N} m_k \frac{\vec{\ri}-\vec{r}_k}{|(\vec{\ri}-\vec{r}_k)|^2},~~ m_k=\frac{M_k}{M}
\label{eq:nlens}
\eeq
where $\sum_{k=1}^N m_k =1$. If there is only one lens ($m_1=1$) and the lens is at the origin, then eq. (\ref{eq:nlens}) reverts to  the single lens equation (\ref{eq:onelens}).

Two-dimensional vectors and complex numbers are closely related, Witt (1990) first
demonstrated that the above equation can be cast 
into a complex form by direct substitutions of the vectors by complex numbers:
\beq
\zs = z - \sum_{k=1}^{N} m_k \frac{z-z_k}{(z-z_k)(\bar{z}-\bar{z}_k)}
= z - \sum_{k=1}^{N} \frac{m_k}{\bar{z}-\bar{z}_k}
\label{eq:complex}
\eeq
where $z=x + y\,i$, $z_k=x_k + y_k\,i$, and $\zs=\xs+ \ys\,i$ (where
$i$ is the imaginary unit).

We can take the conjugate of eq. (\ref{eq:complex}) and obtain an
expression for $\bar{z}$. Substituting this back into
eq. (\ref{eq:complex}), we obtain a complex polynomial of degree $N^2+1$.
This immediately shows that even a binary lens equation cannot be solved
analytically since it is a fifth-order polynomial \footnote{In classical mechanics, the two-body problem can be solved
analytically, but not the three-body problem.}.

The magnification is related to the determinant of the Jacobian of the mapping from the source plane to
the lens plane: $(\xs, \ys) \rightarrow (x,y)$. In the complex form,
this is (\cite{Wit90}; see also the solution to Problem 8):
\beq
\mu = \frac{1}{J}, ~~~ J=\frac{\partial(\xs, \ys)}{\partial(x, y)} = 1 - \frac{\partial \zs}{\partial \bar{z}}
\overline{\frac{\partial \zs}{\partial \bar{z}}}.
\eeq
Notice that the Jacobian can be equal to zero implying
a (point) source will be infinitely magnified. The image positions satisfying $J=0$ form continuous
{\it critical curves}, which are mapped into {\it caustics} in the source plane. Of
course, stars are not point-like, they have finite sizes. The finite
source size of a star smoothes out the singularity. As a result, the magnification remains finite.

\begin{figure}[!ht]
\hspace{-0.5cm}
\includegraphics[width=.5\textwidth]{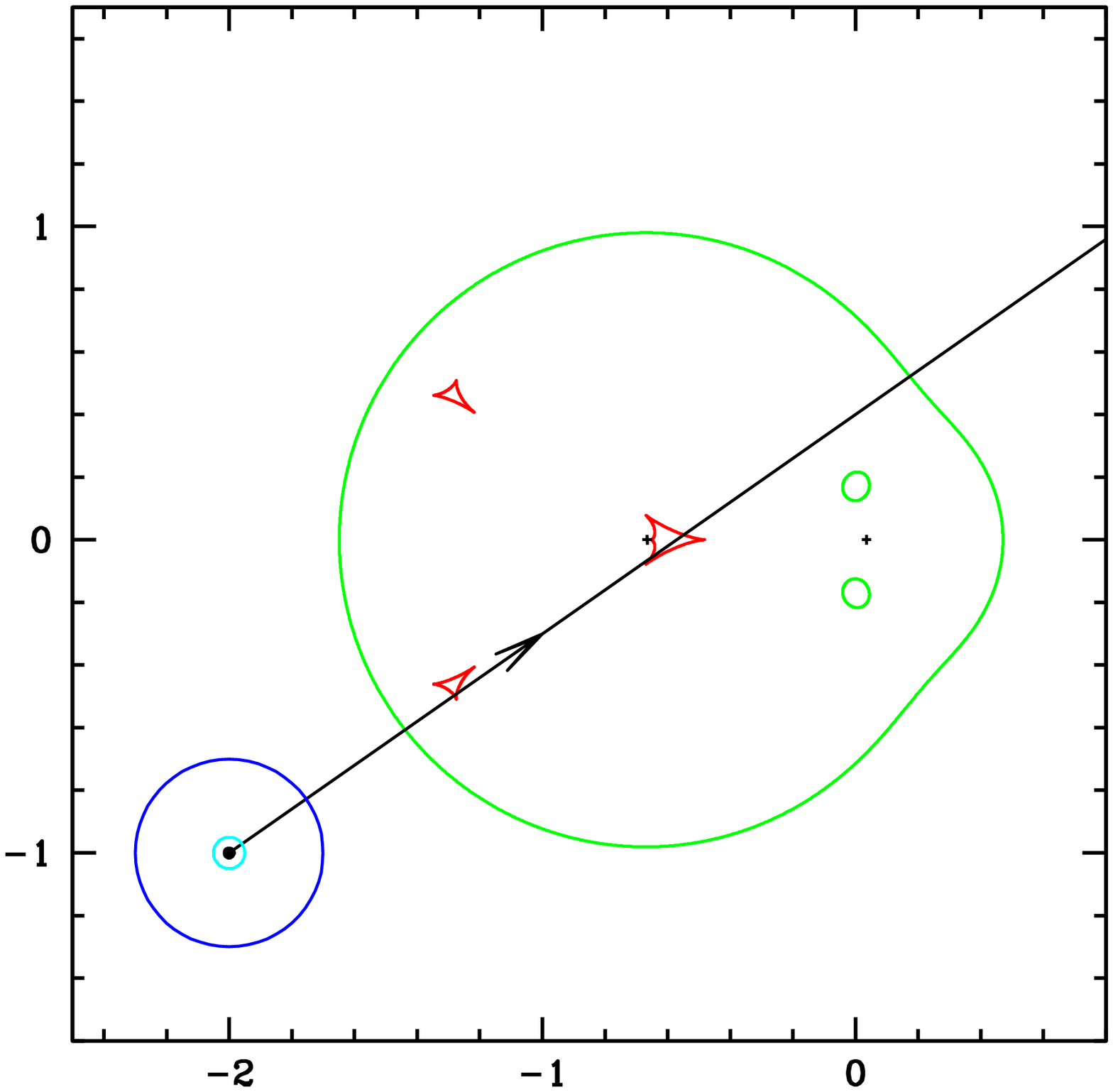} \hfil
\includegraphics[width=.5\textwidth]{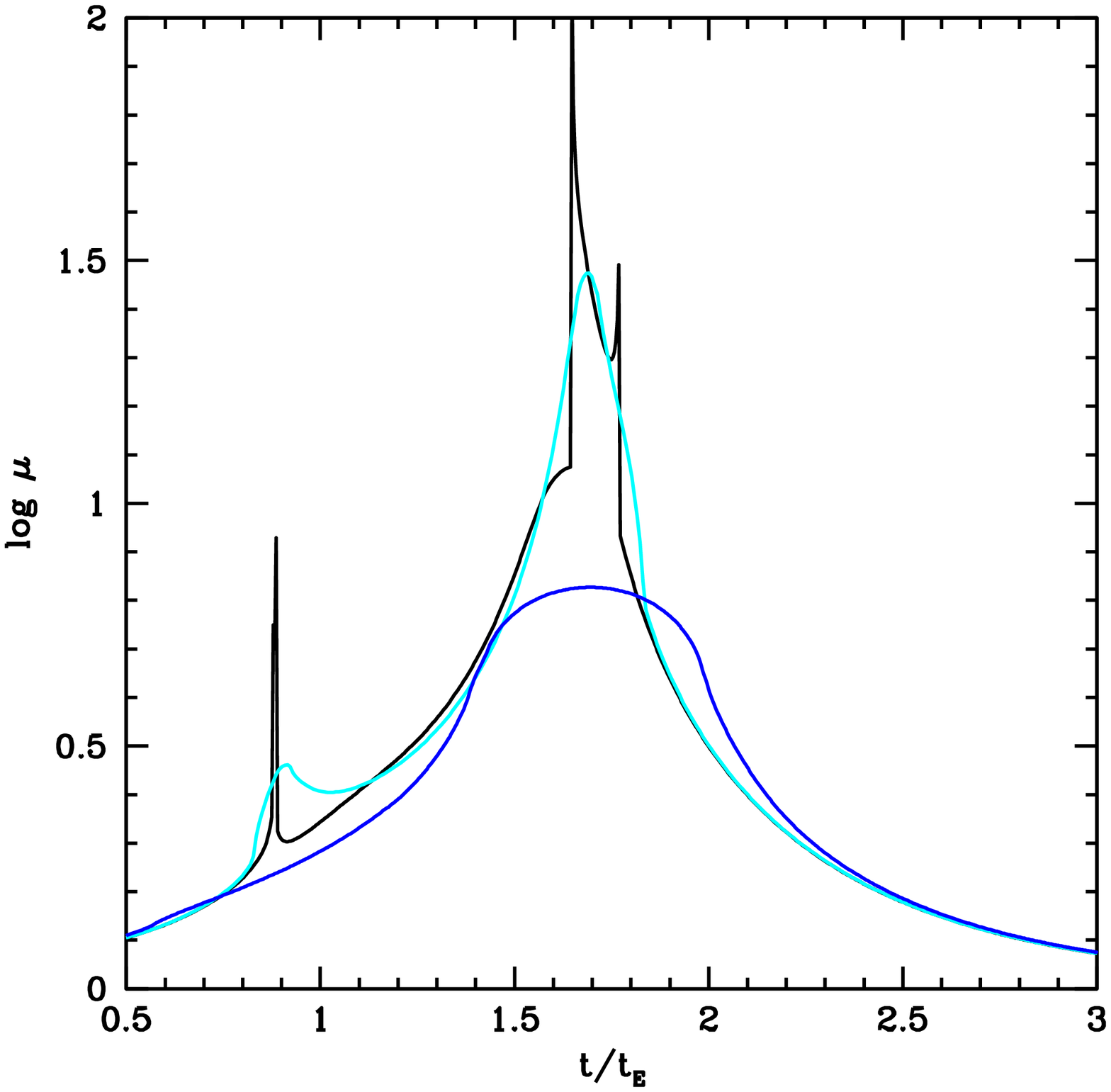}
\vspace{-0.5cm}
\caption{
{\it Left panel}: Caustics (red curves) and critical curves (green curves) for a binary lens. The
lenses (indicated by two `+' signs) are located at $z_1=(-0.665, 0)$ and $z_2=(0.035, 0)$ with mass $m_1=0.95$ and
and $m_2=1-m_1=0.05$ respectively. The black line shows the trajectory
for three source sizes, $\rs/\rE=0, 0.05, 0.3$, indicated by the cyan
and blue circles and a dot (for a point source). The trajectory starts
at $(-2, -1)$ with a slope of 0.7. {\it Right panel}: Corresponding light curves for the three source sizes
along the trajectory in the left panel. The time is
normalised to the Einstein radius crossing time, $\tE$, and $t=0$
corresponds to the starting position. Notice that as the
source size increases, the lensing magnification amplitude decreases. 
}
\label{fig:binary_lc}
\end{figure}

For $N$-point lenses, from the complex lens equation (\ref{eq:complex}),
we have
\beq
\frac{\partial\zs}{\partial \bar{z}} = \sum_{k=1}^{N}
\frac{m_k}{(\bar{z}-\bar{z}_k)^2}, ~~~
J = 1 - \Big|\sum_{k=1}^{N} \frac{m_k}{(\bar{z}-\bar{z}_k)^2}\Big|^2.
\label{eq:Jacobian}
\eeq
It follows that the critical curves are given by
\beq
\Big|\sum_{k=1}^{N} \frac{m_k}{(\bar{z}-\bar{z}_k)^2}\Big|^2 = 1.
\eeq
The sum in the above equation must be on a unit circle, and the solution can be cast in a parametric form
\beq
\sum_{k=1}^{N} \frac{m_k}{(z-z_k)^2} = {\rm e}^{i \Phi},
\eeq
where $0 \le \Phi < 2 \pi$ is a parameter. The above equation is a complex polynomial of degree of $2 N$ with
respect to $z$. For each $\Phi$, there are at most $2N$ distinct solutions. As we vary $\Phi$ continuously, the
solutions trace out at most $2 N$ continuous critical curves (critical
curves of different solutions may join with each other
smoothly). In practise, we can solve the equation for one $\Phi$ value, and then use the
Newton-Raphson method to find the solutions for other values of $\Phi$.

For a single point lens, if we take $z_1=0$, and $m_1=1$, we find that
the critical curve is the Einstein ring ($|z|=1$), which is mapped 
into a degenerate caustic point at the origin ($\zs=0$). However, for binary or
multiple lenses, the critical curves and caustics are much more complex.
The left panel in Fig. \ref{fig:binary_lc} illustrates the critical curves and caustics
for a binary lens with $m_1=0.95$ and $m_2=0.05$ and separation of 0.7
(in units of the Einstein radius for the total mass). In this case, there are three
separate, continuous critical curves which are mapped into three caustics.

For a point source, the complex polynomial can be easily
solved numerically (e.g. using the {\tt zroots} routine in
\cite{Pre92}). However, for a source with finite size, the existence of
singularities makes the integration time-consuming (see section \ref{sec:open}).
The right panel in Fig. \ref{fig:binary_lc} shows the light
curves for three source sizes along the trajectory indicated by the
green straight line. As the source size increases, the lensing
magnification amplitude decreases, and the number of peaks differ for
different source sizes.


\section{Optical depth and event rates \label{sec:tau}}

So far we have derived the lens equation and light curve for
microlensing by a single star. In reality, hundreds of millions of
stars are monitored, and $\approx 800$ unique microlensing events
are discovered each year. Clearly we need some statistical quantities to describe microlensing experiments.
For this, we need two key concepts: optical depth and event rate.

\subsection{Optical depth}

The optical depth (lensing probability) is the probability that a given source 
falls into the Einstein radius of any lensing star along the line of sight. Thus the optical depth can be expressed as
\begin{equation}
\tau = \int_0^\Ds n(\Dd) \left( \pi \rE^2\right) d\Dd,
\end{equation}
which is an integral of the product of the number density of lenses, the lensing 
cross-section ($=\pi \rE^2$) and the differential path ($d\Dd$). 

Alternatively, the optical depth can be viewed as the fraction of sky covered by the angular areas of all the
lenses, which yields another expression
\begin{equation}
\tau = \frac{1}{4\pi} \int_0^\Ds \left[n(\Dd) 4\pi \Dd^2 d\Dd\right] \left( \pi \thetaE^2\right),
\end{equation}
where the term in the [~] gives the numbers of lenses in a spherical
shell with radius $\Dd$ to $\Dd+d\Dd$, $\pi\thetaE^2$ is the angular
area covered by a single lens, and the term in the denominator
is the total solid angle over all the sky ($4 \pi$).

If all the lenses have the same mass $M$, then $n(\Dd) = \rho(\Dd)/M$, $\pi
\rE^2 \propto M$, and the lens mass drops out in $n(\Dd) \pi \rE^2$.
Therefore the optical depth depends on the total mass density along the line
of sight, but not on the mass function.

Let us consider a simple model where the density is constant
along the line of sight, $\rho(\Dd) = \rho_0$. Integrating along the line of sight one finds
\beq
\tau = \frac{2 \pi G \rho_0}{3c^2} \Ds^2 = \frac{1}{2 c^2} \frac{G \rho_0 4\pi
  \Ds^3/3}{\Ds}
= \frac{1}{2 c^2} \frac{G M(<\Ds)}{\Ds} = \frac{V^2}{2 c^2},
\eeq
where $M(<\Ds)$ is the total mass enclosed within the sphere of radius
$\Ds$ and the circular velocity is given by $V^2 = GM(<\Ds)/\Ds$.

For the Milky Way, $V\approx 220\kms$, $\tau \approx 2.6\times
10^{-7}$. The low optical depth means millions of stars have to be monitored to have a realistic yield of microlensing events, and thus one needs to observe dense stellar fields, which in turn means accurate crowded field photometry is essential (see the talk and workshop material by P. Wozniak on difference image
analysis\footnote{{\tt pos.sissa.it//archive/conferences/054/003/GMC8\_003.pdf; www.jodrellbank.manchester.ac.uk/\~\,m2c/dia\_exercises/}}.).

\subsection{Event rate}

The optical depth indicates the probability of a given star that is within
the Einstein radii of the lenses {\it at any given instant}. As such, the optical
depth is a static concept. We are obviously interested in knowing the
event rate (a dynamic concept), i.e.,
the number of (new) microlensing events per unit time for a given number of
monitored stars, $N_\star$.

To calculate the event rate, it is easier to imagine the lenses are moving
in a static stellar source background. For simplicity, let us assume all the lenses move with the same velocity
of $\vt$. The new area swept out by each lens in the time interval $dt$ is equal to the product of the diameter of the Einstein
ring and the distance travelled  $\vt\,dt$, $dA= 2 \rE \times \vt dt = 2 \rE^2 dt /\tE$.
The probability of a source becoming a new microlensing event is given by
\beq
d\tau = \int_0^{\Ds} n(\Dd) dA d\Dd 
   =\int_0^{\Ds} n(\Dd) \left(\frac{2 \rE^2}{\tE}\right) dt d\Dd 
\eeq
The total number of new events is $N_\star d\tau$, and thus the event rate is given by
\beq
\Gamma = \frac{d(N_\star\tau)}{dt}  
   =N_\star \int_0^{\Ds} n(\Dd) \left( \frac{2}{\pi \tE} \cdot {\pi \rE^2}\right)\, d\Dd 
=\frac{2N_\star}{\pi}\int \frac{d\tau}{\tE}. 
\eeq
If, for simplicity, we assume all the Einstein radius crossing times are
identical, then we have
\beq
\Gamma \approx \frac{2 N_\star}{\pi} \frac{\tau}{\tE}.
\label{eq:gamma}
\eeq
We make several remarks about the event rate:
\begin{enumerate}
\item[(1)] If we take $\tE=19 \day$ (roughly equal to the median of the observed timescales), then we
have
\beq
\Gamma \approx \frac{2 N_\star}{\pi} \frac{\tau}{\tE}
= 1200 \yr^{-1} \frac{N_\star}{10^8}\frac{\tau}{10^{-6}} \left(\frac{\tE}{19 \day}\right)^{-1}
\eeq
For OGLE-III, about $2\times 10^8$ stars are monitored (see Udalski's contribution),
so the total number of events lenses we expect per year is
$\Gamma \sim 2400$ if $\tau \sim 10^{-6}$, which is a factor of four of the observed rate (indicating the detection efficiency may be of the order of 30\%).
\item[(2)] While the optical depth does not depend on the mass function, the
  event rate does because of $\tE (\propto M^{1/2})$ in the denominator of eq. (\ref{eq:gamma}). The timescale distribution can
  be used to probe the kinematics and mass function of lenses in the Milky Way.
\item[(3)] The lenses and sources have velocity distributions, one must account for them
when realistic event rates are needed. Furthermore, the source distance
is unknown, and so in general we need to average over the source distance (for example calculations, see \cite{Gri91, KP94}).
\end{enumerate}

\section{Summary \label{sec:open}}

In this introduction, we derived the lens equation, and obtained
the image positions and magnifications for a point lens. We also
discussed the statistical measures for microlensing experiments, and
estimated the order of magnitudes for various quantities. An interested reader should now be
armed with a basic knowledge of microlensing and be prepared to read the
review articles and start to do research on gravitational microlensing 
(or even try to solve the problem set below). 

Since the discovery of first microlensing events in the early 1990's,
enormous achievements have been made in the field. However, challenges
and opportunities remain.
\begin{enumerate}
\item[(1)] Undoubtedly the highlight of gravitational microlensing in the last
  few years has been the discovery of extrasolar planets
  (\cite{Bon04, Uda05, Bea06, Gou06, Ben08}).
  Microlensing has much to offer in this area since it probes a
  different part of the parameter space, and provides an important test
  of the core accretion theory of planet formation. Several White Papers 
  (\cite{Gou07, Ben07, Dom08, Bea08}) set out strategies with ambitious
  milestones in the next fifteen years, from improvement of the current
  survey plus followup mode of discovery (with an automated algorithm to
  identify the ``anomalies'' in real-time) in the near term, to a wide-field network from the ground in the next 5-10
  years, and eventually a telescope in space in the next 10-15
  years. Combined with the stellar transit mission {\it
  Kepler} (to be launched in 2009), microlensing will be able to
  provide the complete census of Earth-mass (and lower) planets at
  virtually all the separations.

  Technically, it is still challenging to calculate the light curves for
  sources with finite size since we need to integrate over the
  singularities of caustics. This is particularly important for the discovery of extrasolar
  planets when a source transits the small caustics induced by the planet(s).
  The problem becomes even worse with the discovery of multiple planets
  (\cite{Gau08}) due to the higher complexity of the lens equation and
  the increased number of parameters: how do we search the high dimensional parameter space efficiently?

  Are there hidden multiple planetary light curves in the database that
  are not yet identified due to their complex shapes?
\item[(2)] Microlensing surveys over the last fifteen years
  have accumulated tens of TB of data. This tremendous database
  has not been exploited to its fullest potential.
  
  For example, the surveys yielded many high-quality colour-magnitude
  diagrams of stellar populations,  proper motions of millions of
  stars, and in the future the optical depth maps. All these can be used
  to provide important and independent probes of the structure of the Milky Way.

  Despite promising earlier attempts (e.g. \cite{Ng96}; \cite{Rat07a, Rat07b}; \cite{KP94, Zha96}),
  microlensing has under-delivered in this area. 
  For example, while we have discovered several thousands of microlensing events over
  the last 15 years, only a small fraction has been used for statistical
  analyses of optical depths. We need to remedy the situation urgently.
\item[(3)] High-magnification events are great targets-of-opportunity
for high signal-to-noise ratio spectroscopic observations to study
stellar atmospheres for bulge stars. Attempts so far already yielded
interesting results (e.g. \cite{Len96, Thu06, Coh08}). We need to explore this more systematically.
\item[(4)] For mathematically-gifted students (or mathematicians),
  gravitational microlensing provides an interesting problem. While the binary lens equation is no longer
  analytical, there is, nevertheless, an analytical relation on the
  minimum magnification for five-image configurations (\cite{WM95, Rhi97}). There is also a
  degeneracy found by Dominik (1999) between close and wide separation binaries, which
  was later explored in much greater detail by An (2005). Are there any
  other symmetries, perhaps even for multiple lens systems?

  The number of critical curves for $N$-point lenses cannot exceed $2N$
  (see section \ref{sec:binary}). The upper bound of the number of
  images for $N$-point lenses also has linear dependence on $N$ (see
  Problem 1). These two are related mathematically (see
  \cite{Rhi01,Rhi03, An06}). Is there a more geometric (topological) way
  of understanding this?
\end{enumerate}

\vspace{0.5cm}
I thank Eamonn Kerins, Szymon Koz{\l}owski, Nick Rattenbury and particularly Matthew Penny for a careful reading of the manuscript.

\section*{Problems}

\begin{enumerate}
\item
a) There are $N$-point lenses in a lens plane. For a source very far
away from all the lenses, how many images are there? What are their parities?

b) What is the achievable maximum number of images for $N$-point lenses
and an arbitrary source position?
\item A uniform source star is perfectly aligned with an observer and a point
lens. Its physical radius normalised to the Einstein radius is $\rhos$. 
\begin{itemize}
\item[a)] What is the resulting image configuration?
\item[b)] What is the magnification for the finite source?
\item[c)] Estimate the maximum magnification that can be achieved for a
source star in the Galactic bulge.
\item[d)] Derive the expression for magnification when the source is not perfectly aligned with the lens.
\end{itemize}
\item Show that the total magnification for a point lens is
always larger than one. How can this be reconciled with energy conservation?
(see Jaroszy\'nski \& Paczynski 1996, AcA, 46, 361).
\item The density distribution in the plane of the Galactic disk can be
modelled as an exponential
$$\rho(R) = \rho_0 \exp(-(R-R_0)/R_d),$$
where $\rho_0$ is the density in the solar neighbourhood, $R_0$ is the
distance from the Sun to the Galactic centre, $R_d$ is the disk scale length, and $R$ is the distance 
from the Galactic centre to the lens.
\begin{itemize}
\item[a)] Find the optical depth $\tau$ for a source at the Galactic centre ($R=0$).
\item[b)] If $\rho_0=0.1 M_\odot\,{\rm pc}^{-3}$, $R_0=8$\,kpc,
$R_d=3$\,kpc, what is the value of $\tau$?
\end{itemize}
\item
Consider a simple model: all lensing objects have the same mass $M$,
the same three-dimensional velocity $V$, and their velocity vector directions have
an isotropic distribution. The source located at the distance is stationary,
and the number density of lensing objects is uniform between the observer and the source. 
Derive the timescale probability distribution.

Now assume the lenses follow a Maxwellian distribution with
a one-dimensional velocity dispersion $\sigma$. Derive
the timescale probability distribution. Show that it follows a power-law behaviour for both very short and very long timescales.
\item A distance source is lensed by a point deflector with mass $M$.
The light signals emitted by the source will be received by an observer at different times
for the two images due to the difference in the trajectory and
gravitational potential experienced.
The time delay is of the order of $r_{\rm sch}/c$, where $r_{\rm sch}$ is the Schwarzschild
radius. Is this observable for a $M=1M_\odot$ lens?
\item A background star stationary at the origin is microlensed by a lens moving from $-\infty$ to $\infty$. Show that the centre of light of the two images traces out an ellipse.
\item Show that
\begin{itemize}
\item[a)] In the complex notation, the Jacobian is given by eq. (\ref{eq:Jacobian}).
\item[b)] For any positive-parity image produced by $N$-point lenses,
  the magnification is always larger than or equal to one.
\item[c)] Find the number and positions of images where their
  magnifications are identical to unity for a binary lens.
\item[d)] What is the maximum number images with unity magnification for
  $N$-point lenses?
\end{itemize}
\end{enumerate}

\newpage
\section*{Solutions}

\begin{enumerate}
\item
\begin{itemize}
\item[a)]
There are $N+1$ images, one bright image close to the source, and $N$
fainter images with one image close to each lens. The image close to the
source has positive parity, all the others have negative parities.
\item[b)]
$5(N-1)$ for $N \ge 2$. This is a difficult problem. See Rhie
  S. H. 2003, arXiv:astro-ph/0305166; Khavinson, D. and Neumann G. 2006,
  Proceedings Of the American Mathematical Society, 134, 1077 (arXiv:math/04011088).
\end{itemize}
\item 
\begin{itemize}
\item[a)]
The image configuration is an annulus. The outer radius is
given by the $r_2=(\sqrt{\rhos^2+4}+\rhos)/2$ while the inner radius is given by $r_1=(\sqrt{\rhos^2+4}-\rhos)/2$.
\item[b)]
The area covered by the image is $A=\pi(r_2^2-r_1^2) = \pi \rhos \sqrt{\rhos^2+4}$. Thus
the magnification is $\mu=A/(\pi \rhos^2) = \sqrt{\rhos^2+4}/\rhos$. In
particular, when $\rhos \rightarrow 0$, $\mu \rightarrow 2/\rhos$.
\item[c)]
For a solar mass lens in the Galactic centre ($\Ds=8\AU$), the maximum
Einstein radius is achieved when $\Dd=\Ds/2$, $\rE=6.0 \times 10^{13}\cm$. The faintest and smallest
stars we can see in the bulge have radii similar to a solar-type star, $r_\star \approx R_\odot=7 \times 10^{10}\cm$,
$\rhos=r_\star/r_E = 1.2 \times 10^{-3}$. The
maximum magnification is of the order of $\approx 2/\rhos \approx 1700$.
\item[d)] See Witt \& Mao 1994, ApJ, 430, 505
\end{itemize}
\item Clearly $\mu>1$. See Jaroszy\'nski \& Paczynski 1996, AcA, 46, 361
for discussions about the energy conservation.
\item 
The optical depth is given by
\beq
\tau = \int_0^\Ds n(\Dd) ~\left( \pi \rE^2\right) ~ d\Dd
     = \int_0^\Ds \frac{\rho(\Dd)}{M} ~ \frac{4\pi G M}{c^2}
     \frac{\Dd\Dds}{\Ds} ~ d\Dd.
\eeq
For microlensing in the Galactic plane, $\Dds=R, \Ds=R_0, \Dd = R_0-R$,
and thus we have 
\beq
\tau = \int_0^{R_0} \rho_0 ~ {\rm e}^{\Dd/R_d} ~ \frac{4 \pi G}{c^2} \frac{\Dd(R_0-\Dd)}{R_0}~ d\Dd
\eeq
Performing the above integral, we find that
\beq
\tau = \frac{4 G \pi \rho_0 R_0^2}{c^2}~ y^{-3} [2+y+{\rm e}^y (-2 + y)], ~~ y = R_0/R_d
\eeq
For the given numbers, we find $4 \pi G\rho_0 R_0^2/c^2 = 3.86 \times
10^{-6}$, and $y=R_0/R_d=2.67$, we have
\beq
\tau  \approx 2.9 \times 10^{-6}.
\eeq
\item For step by step derivations, see Mao \& Paczy\'nski (1996,
  ApJ, 473, 57). Notice that the observed event timescale distribution
  does not follow the power-laws due to detection efficiency.
\item The time delay is of the order of tens of micro-seconds for a solar
mass lens, and is very difficult to observe.
\item See Hog et al. (1995, A\&A, 294, 287), Miyamoto \& Yoshi (1995, AJ, 110, 1427),  Walker (1995, ApJ, 453, 37)	
\item
\begin{itemize}
\item[a)] Using complex notation, we have $x = \frac{1}{2}(z+\zbar), y = \frac{1}{2 i}(z - \zbar)$.
Thus 
\beq
\frac{\partial{\zs}}{\partial{z}} = \frac{\partial{\zs}}{\partial{x}} \frac{\partial{x}}{\partial{z}}  
+ \frac{\partial{\zs}}{\partial{y}} \frac{\partial{y}}{\partial{z}}  
= \frac{1}{2} \left(\frac{\partial{\zs}}{\partial{x}}
- i \frac{\partial{\zs}}{\partial{y}}\right)
=\frac{1}{2} \left(\frac{\partial{\xs}}{\partial{x}}
+i \frac{\partial{\ys}}{\partial{x}}
- i \frac{\partial{\xs}}{\partial{y}}
+ \frac{\partial{\ys}}{\partial{y}}\right).
\label{eq:zsz}
\eeq
Similarly
\beq
\frac{\partial{\zs}}{\partial{\zbar}} = \frac{\partial{\zs}}{\partial{x}} \frac{\partial{x}}{\partial{\zbar}}  
+ \frac{\partial{\zs}}{\partial{y}} \frac{\partial{y}}{\partial{\zbar}}  
= \frac{1}{2} \left(\frac{\partial{\zs}}{\partial{x}}
+ i \frac{\partial{\zs}}{\partial{y}}\right)
= \frac{1}{2} \left(
\frac{\partial{\xs}}{\partial{x}}
+ i \frac{\partial{\ys}}{\partial{x}}
+ i \frac{\partial{\xs}}{\partial{y}}
- \frac{\partial{\ys}}{\partial{y}}
\right).
\label{eq:zszc}
\eeq
However, from the lens equation (\ref{eq:complex}),
$\partial{\zs}/\partial{z}=1$, and thus comparing the real and imaginary
parts in eq. (\ref{eq:zsz}), we have
\beq
\frac{1}{2}\left(
\frac{\partial{\xs}}{\partial{x}}
+
\frac{\partial{\ys}}{\partial{y}}
\right)
= 1, ~~~
\frac{\partial{\xs}}{\partial{y}}
=
\frac{\partial{\ys}}{\partial{x}}.
\eeq
Substituting the second expression into eq. (\ref{eq:zszc}), we find
\beq
\frac{\partial{\zs}}{\partial{\zbar}} =\frac{1}{2} \left(
\frac{\partial{\xs}}{\partial{x}} 
-
\frac{\partial{\ys}}{\partial{y}}\right) + i 
\frac{\partial{\ys}}{\partial{x}}.
\eeq
Combined with
\beq
\frac{\partial{\zs}}{\partial{z}} =
\frac{1}{2} \left(\frac{\partial{\xs}}{\partial{x}} +\frac{\partial{\ys}}{\partial{y}}\right) = 1
\eeq
we find
\beq
\frac{\partial\xs}{\partial x}
= 1 + \frac{1}{2} \left(\frac{\partial\zs}{\partial\zbar}
+\overline{\frac{\partial\zs}{\partial\zbar}}\right),
\frac{\partial\ys}{\partial y}
= 1 - \frac{1}{2} \left(\frac{\partial\zs}{\partial\zbar}
+\overline{\frac{\partial\zs}{\partial\zbar}}\right),
\frac{\partial{\ys}}{{\partial x}} = \frac{1}{2 i} 
\left(\frac{\partial\zs}{\partial\zbar}
-\overline{\frac{\partial\zs}{\partial\zbar}}\right).
\eeq
Substituting the above equations into the Jacobian
\beq
J=\frac{\partial{\xs}}{\partial{x}}
\frac{\partial{\ys}}{\partial{y}}
- \frac{\partial{\ys}}{\partial{x}}
\frac{\partial{\xs}}{\partial{y}},
\eeq
we recover the required expression.
\item[b)] Since 
\beq
J = 1 - \Big|\sum_{k=1}^{N} \frac{m_k}{(\bar{z}-\bar{z}_k)^2}\Big|^2.
\eeq
For a positive parity image, we must have $1 \ge J>0$, it follows
that $\mu = J^{-1} \geq 1$. The magnification is unity when the sum is equal to zero.
\item[c)] Without losing generality, we put the lenses on the $x$-axis,
  at $\vec{r}_1=(z_0, 0)$ and $\vec{r}_2=(-z_0, 0)$, with masses $m_1$ and
  $m_2$ ($m_1 \ge m_2$), and we have $m_1+m_2 = 1$, $z_0>0$. From part
  b), the condition for unity magnification is
\beq
\frac{m_1}{(\zbar+z_0)^2} + \frac{m_2}{(\zbar-z_0)^2} = 0.
\eeq
This simplifies into a quadratic equation
\beq
\zbar^2 + 2(m_1-m_2) z_0 \zbar + z_0^2 = 0.
\eeq
Clearly we have two solutions
\beq
\zbar_{+,-} = z_0 \left((m_2-m_1) \pm i\sqrt{1-(m_1-m_2)^2}\right).
\eeq
These two are, as expected, symmetric with respect to the $x$-axis.
\item[d)] $2(N-1)$.
\end{itemize}
\end{enumerate}

\end{document}